\documentclass{article}
\usepackage{spconf,amsmath,graphicx}
\usepackage{booktabs}
\usepackage{tabularx}
\usepackage{float}
\usepackage{multirow}
\usepackage{array}
\usepackage{amsmath}

\title{SFUSNet: A Spatial-Frequency domain-based Multi-branch Network for diagnosis of Cervical Lymph Node Lesions in Ultrasound Images}
\name{Yubiao Yue\textsuperscript{2}, Jun Xue\textsuperscript{3}, Haihua Liang\textsuperscript{1}, Bingchun Luo\textsuperscript{4}, Zhenzhang Li\textsuperscript{1,*}\thanks{This work was supported by the NSF of Guangdong Province (No.2022A1515011044,No.2023A1515010885), and the  project of  promoting  research capabilities for key constructted disciplines in Guangdong Province (No.2021ZDJS028). * is corresponding author.}}
\address{\textsuperscript{1}College of Mathematics and Systems Science,\\Guangdong Polytechnic Normal University, Guangzhou, China\\
\textsuperscript{2}School of Biomedical Engineering, Guangzhou Medical University, Guangzhou, China\\
\textsuperscript{3}School of Computer Science and Technology, Anhui University, Anhui, China\\
\textsuperscript{4}School of Computer Science and Technology, Harbin Institute of Technology, Weihai, China}
\begin{document}
%
\maketitle
\begin{abstract}
Booming deep learning has substantially improved the diagnosis for diverse lesions in ultrasound images, but a conspicuous research gap concerning cervical lymph node lesions still remains. The objective of this work is to diagnose cervical lymph node lesions in ultrasound images by leveraging a deep learning model. To this end, we first collected 3392 cervical ultrasound images containing normal lymph nodes, benign lymph node lesions, malignant primary lymph node lesions, and malignant metastatic lymph node lesions. Given that ultrasound images are generated by the reflection and scattering of sound waves across varied bodily tissues, we proposed the Conv-FFT Block. It integrates convolutional operations with the fast Fourier transform to more astutely model the images. Building upon this foundation, we designed a novel architecture, named SFUSNet. SFUSNet not only discerns variances in ultrasound images from the spatial domain but also adeptly captures micro-structural alterations across various lesions in the frequency domain. To ascertain the potential of SFUSNet, we benchmarked it against 12 popular architectures through five-fold cross-validation. The results show that SFUSNet is the state-of-the-art model and can achieve 92.89\% accuracy. Moreover, its average precision, average sensitivity and average specificity for four types of lesions achieve 90.46\%, 89.95\% and 97.49\%, respectively.
\end{abstract}
\section{Introduction}
\label{sec:intro}
Cervical lymph nodes are a key component of the lymphatic system, responsible for the filtration of lymph fluid and playing a central role in immune regulation \cite{1}. Unfortunately, they are precisely the human tissues that are prone to pathological changes. Medically, cervical lymph node lesions are divided into two categories: benign (non-neoplastic lesions) and malignant (neoplastic lesions), among which malignant lesions are further divided into primary and metastatic. Malignant lesions, in particular, are directly linked to fatal diseases such as lymphoma, head and neck cancer, and thyroid cancer, often resulting in the death of patients. Given the differences in etiology, pathological mechanism, and treatment strategy of these lesions, accurate identification and diagnosis of them have become an important task in the medical field\cite{2}. Among various diagnostic techniques, ultrasound images is widely adopted in the evaluation of cervical lymph node lesions due to its non-invasive, real-time and low-priced advantage\cite{3}. However, the interpretation of ultrasound images is highly dependent on experienced radiologists, and there are still subjective differences in interpretation, which may lead to inconsistencies in diagnosis. Based on these facts, exploring more efficient and accurate diagnostic techniques is of great significance.

Recently, more and more studies have demonstrated that compared with artificial diagnosis, deep learning models are more efficient and accurate in ultrasound images diagnosis and even achieve accuracy exceeding experts. For example, Han et al. \cite{4} used 7408 B-mode breast ultrasound images to train GoogLeNet in order to distinguish benign and malignant breast masses. They reported an accuracy of 91\% for their deep learning model. Fujioka et al. \cite{5} used 947 B-mode ultrasound images to train a model built with GoogLeNet and Inception v2, and then tested 48 images of benign masses and 72 images of malignant masses. The results showed that the accuracy of the model reaches 95.8\%, which is equal to or better than that of radiologists. Yang et al. \cite{6} collected 508 thyroid ultrasound images, and then used the pre-trained Resnet18 to analyze the benign and malignant thyroid nodules in the ultrasound images. Experimental results showed that their model was able to improve accuracy, sensitivity, F1- The scores reached 98.4\%, 97.8\% and 95.7\% respectively. Xi et al. \cite{xi2021deep} used pre-trained resnet50 to differentiate benign and malignant solid liver lesions in ultrasound images and reported the their model performance exceeds experts. In terms of lesions diagnosis for cervical lymph node ultrasound images, to our knowledge, only Xia et al. \cite{7} collected 420 ultrasound images containing benign and malignant lesions, and then used ResNet50 and DenseNet161 to classify them. The experimental results show that DenseNet161 performed the best, and its sensitivity, specificity, accuracy and AUC reached 93\%, 88\%, 90\% and 0.90, respectively. Simply put, the work of diagnosing lymph node lesions in cervical ultrasound images through deep learning is still in its infancy and has significant shortcomings. For example, there are the lacks of more detailed classification, sufficient validation data and exploration of model architecture.

In this work, our main contributions can be summarized as follows: (1) We collected 3392 ultrasound images of cervical lymph nodes, including normal lymph nodes, benign lesions, malignant primary lesions, and malignant metastatic lesions. (2) In fact, ultrasound images represent the reflection and diffusion of sound waves by different tissues of the human body. Therefore, different tissues and lesions may exhibit characteristics in the frequency domain that are difficult to identify in the spatial domain. Taking this into account, we proposed an innovative network structure named SFUSNet. The core of SFUSNet is Conv-FFT Block. This block uses identity mapping, convolution operations and fast Fourier transform to effectively capture the spatial and frequency domain information in ultrasound images, thereby achieving more accurate diagnostic results. (3) In order to evaluate the potential of our model, We benchmarked SFUSNet and the remaining 12 mainstream architectures via five-fold cross-validation.

\section{MATERIALS AND METHODOLOGY}
\subsection{Data Source}
\begin{figure}
    \centering
    \includegraphics[width=0.9\columnwidth]{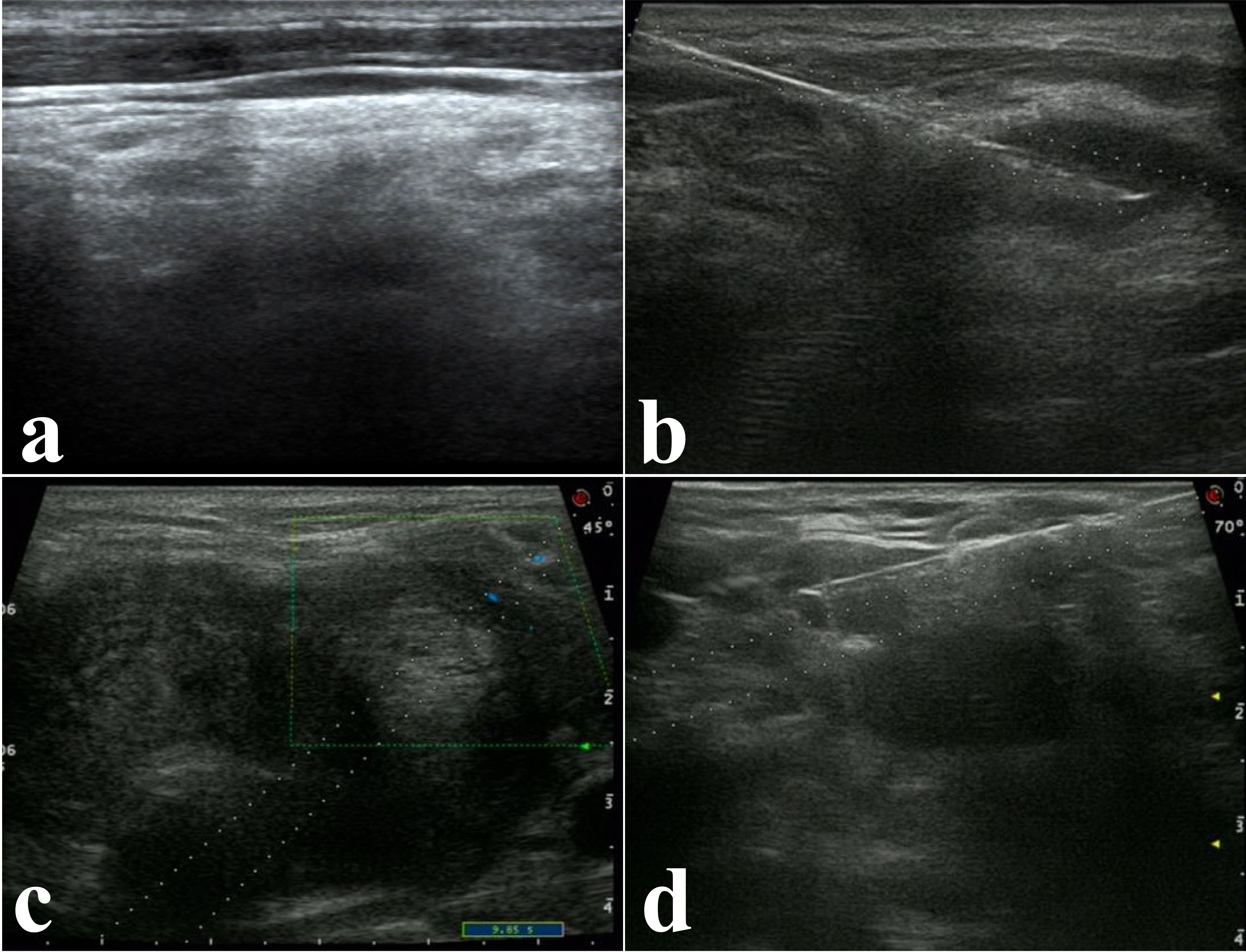}
    \caption{The samples of different lesions. \textbf{a.} Normal; \textbf{b.} Benign; \textbf{c.} Primary; \textbf{d.} Metastatic.}
    \label{fig1}
\end{figure}

The images used in this study were obtained from 480 patients who sought medical care in the radiology department of The Second Affiliated Hospital of Guangzhou Medical University from 2018 to 2021. Two radiologists used SonoScape WIS and Mindray DC-8 to collect 8 to 10 B-mode ultrasound images of cervical lymph nodes from different angles for each patient, and then 2 experts eliminated the images with unqualified quality and angles to obtain 3392 images. The experts then divided these ultrasound images into four categories according to the corresponding pathological findings: normal cervical lymph nodes(Normal, 1217 images), benign cervical lymph nodes(Benign, 601 images), malignant primary cervical lymph nodes(Primary, 236 images), and malignant metastatic lymph nodes(Metastatic, 1338 images). All images are in BMP format, and the original size is 480×350×3. The samples of various lesions are shown in Fig. \ref{fig1}. The research protocol of this study was approved by the Review Committee of the Second Affiliated Hospital of Guangzhou Medical University (Approval No: 2022-hs-78-02). The Helsinki Declaration's tenets were scrupulously followed throughout the study to respect the rights, privacy, and anonymity of the subjects. The use of de-identified data precluded the use of informed consent. 

\subsection{Network Architecture}
\begin{figure*}
    \centering
    \includegraphics[width=0.9\textwidth]{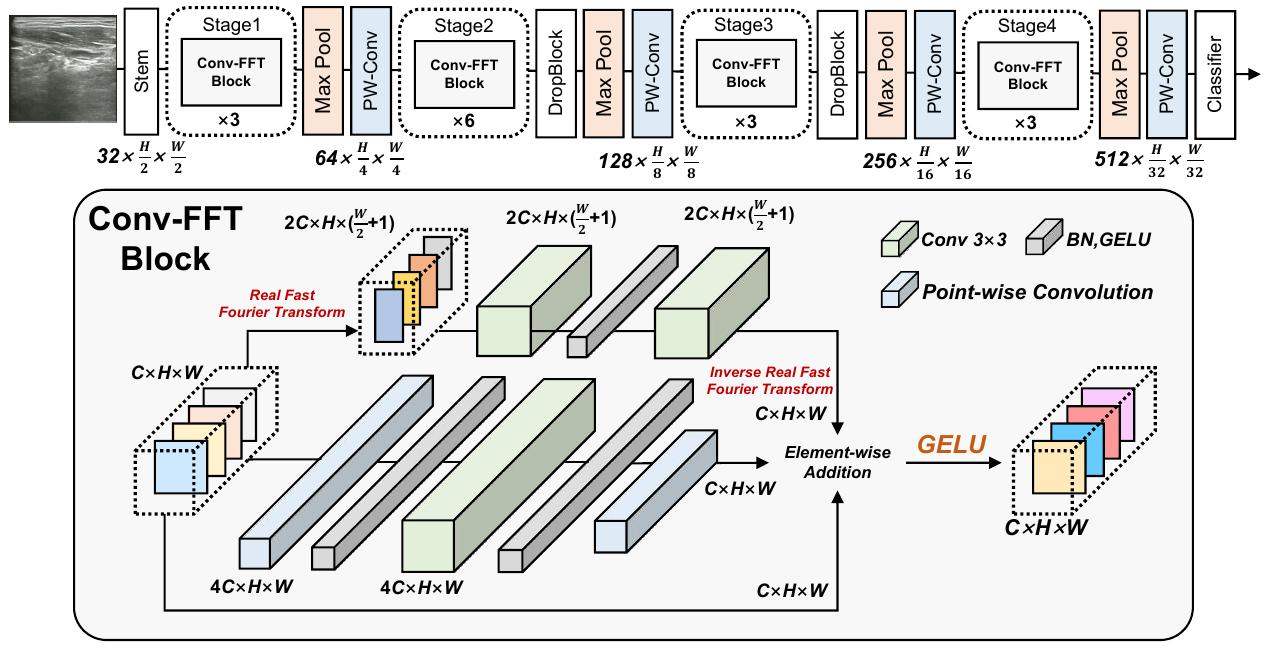}
    \caption{The overall architecture of SFUSNet. BN means Batch Normalization.}
    \label{fig2}
\end{figure*}

The overall framework of the model is shown in Fig. \ref{fig2}. First, the stem block composed of conv3×3, BatchNormal and GELU performs preliminary feature extraction on the input image to capture the texture and edge details and changes dimension from 3×224×224 to 32×112×112. Next, four stages modules composed of Conv-FFT Block further extracts deep features. Here, the internal components of stages 1 to 4 are composed of [3,6,3,3] Conv-FFT Blocks respectively. Between every two stages, the max pooling layer and Point-wise convolution(PW-Conv) are used to downsample the feature map and expand the number of channels by two times. Additionally, DropBlock\cite{8} is added after the stage 2 and 3, which can randomly discard continuous regions in the feature map to prevent model overfitting and improve model generalization. At the end of the model, a typical classification head is used to predict the final result of the image.

Throughout the entire network architecture, Conv-FFT Block is the core component for accurate diagnosis of ultrasound images, and has a unique three-branch design. The detailed structure is shown in Fig. \ref{fig2}. From the details, the central branch is sequentially constructed by Point-wise convolution, Conv3×3 and Point-wise convolution. Its purpose is to extract features from the ultrasound image in the spatial domain. In the upper branch, the feature maps first are transformed from spatial domain to frequency domain via Real Fast Fourier Transform. Next, two consecutive Conv3×3 are used to accurately model the ultrasound image in the frequency domain and then Inverse Real Fast Fourier Transform is responsible for transforming the feature maps from frequency domain to frequency spatial domain. The lower branch adopts the strategy of identity mapping, which not only promotes the reuse efficiency of the network for features, but also effectively resists the challenge of gradient disappearance. At the end of Conv-FFT, the element-wise addition operation is used to fuse the features from the three branches, and Utilizing GELU function activates the fused features to further strengthen the module's nonlinear representation of the input features.

\subsection{Evaluation Metrics}
\label{sec:em}
Here, based on the characteristics of medical images, we used accuracy, precision, sensitivity and specificity as evaluation metrics for the model. The high or low values of these four metrics can reflect the potential of the model in ultrasound images diagnosis. Equation \ref{equ1} to \ref{equ4} show the calculation process of the four metrics.
\begin{equation}
\label{equ1}
Accuracy = \frac{{TP + TN}}{{TP + TN + FP + FN}}
\end{equation}
\begin{equation}
\label{equ2}
Sensitivity = \frac{{TP}}{{TP + FN}}
\end{equation}
\begin{equation}
\label{equ3}
Specificity = \frac{{TN}}{{TN + FP}}
\end{equation}
\begin{equation}
\label{equ4}
Precision = \frac{{TP}}{{TP + FP}}
\end{equation}

Here, TP, TN, FP and FN are the numbers of true positives, true negatives, false positives and false negatives.

\subsection{Implementation Details}
In this work, the average value ± standard deviation of the five-fold cross-validation results is used to thoroughly evaluate the potential of our network. Before training the network, we resized all images to 224×224×3. Subsequently, the images were normalized and standardized using the mean [0.2394,0.2421,0.2381] and standard deviation [0.2173,0.2177,0.2155] for the three channels. During the training process, we employed the Adam optimizer with a 0.001 initial learning rate, $\beta1$ of 0.9, $\beta2$ of 0.999, and weight decay of 1e-4 and Cross-Entropy Loss to optimize the model parameters. To train the network, we utilized the PyTorch framework. We trained the model for 150 epochs and used a batch size of 128. The training was conducted on a computer with Ubuntu 22.04 operating system and an NVIDIA GeForce RTX 4090 GPU.

\section{RESULTS AND DISCUSSION}
To demonstrate the potential and effectiveness of our work, we selected six series of well-known networks, including both convolutional neural networks and vision transformers, as reference models. These models are DenseNet\cite{9}, ResNet\cite{10}, ConvNext\cite{11}, Swin Transformer\cite{12} (SwinT for short), Vision Transformer\cite{13} (ViT for short), and Swin Transformer V2 \cite{14}(SwinT V2 for short). All models were trained strictly according to the requirements specified in the implementation details, and unified metrics were used for evaluation.

Table \ref{tab1} shows the accuracy of SFUSNet and other reference models in detail. Experimental results show that our model achieves the best accuracy. Meanwhile, DenseNet121, ResNet101, ConvNext-large, SwinT-small, ViT-Base and SwinT V2-small achieved the best results in their respective series. Compared with these six networks, the accuracy of SFUSNet is increased by 2.30\%, 6.39\%, 13.06\%, 16.27\%, 17.3\% and 18.69\%, respectively. In addition, the standard deviation results show that SFUSNet has better stability. 

{
\renewcommand{\arraystretch}{1}
\setlength{\arrayrulewidth}{1pt}
\begin{table}[]
\centering
\caption{Comparison of Accuracy between SFUSNet and other networks.}
\label{tab1}
\vspace{0.2cm}
\begin{tabular}{{>{\centering\arraybackslash}p{3.5cm}>{\centering\arraybackslash}p{3.5cm}}}
\toprule
\textbf{Model} & \textbf{Accuracy} \\ \midrule 
DenseNet161 \cite{9}    & 90.53\%(±1.11)    \\ 
DenseNet121 \cite{9}    & 90.59\%(±1.53)    \\ 
ResNet101 \cite{10}     & 86.50\%(±1.04)    \\ 
ResNet152 \cite{10}    & 83.90\%(±0.86)    \\ 
ConvNext-Base \cite{11}  & 79.54\%(±0.88)    \\ 
ConvNext-Large \cite{11} & 79.83\%(±0.67)    \\ 
SwinT-Base \cite{12}     & 76.30\%(±0.54)    \\ 
SwinT-Small \cite{12}    & 76.62\%(±0.59)    \\ 
ViT-Base \cite{13}       & 77.27\%(±0.71)    \\ 
ViT-Large  \cite{13}     & 75.59\%(±0.50)    \\ 
SwinT V2-Base \cite{14}  & 73.64\%(±1.13)    \\ 
SwinT V2-Small \cite{14} & 74.20\%(±0.60)    \\ \midrule
\textbf{SFUSNet(Ours)}       & \textbf{92.89\%(±0.42)}    \\ \bottomrule
\end{tabular}
\end{table}

In order to more fully demonstrate the performance of SFUSNet, we further divided the validation set into Benign and non-Benign, Normal and non-Normal, Primary and non-Primary and Metastatic and non-Metastatic. According to Equation \ref{equ2} to \ref{equ4}, the sensitivity, specificity and precision of SFUSNet for each type of lesion are calculated. Meanwhile, the best models in each series are selected for comparison. Table \ref{tab2} shows in detail the diagnostic capabilities of each network for each type of disease. By observing the presicion, sensitivity and specificity of each type of lesion, it is not difficult to find that SFUSNet is the most outstanding. We additionally calculated the mean values of the SFUSNet metrics for the four diseases. Its average precision, average sensitivity and average specificity achieve respectively 90.46\%, 89.95\% and 97.49\%. 

It is worth noting that by observing these results, some interesting phenomenons can be found: Transformer-based architectures are less effective at diagnosing cervical lymph lesions in ultrasound image lesions and Convolution-based architectures have better advantages than Transformer-based architectures. We believe this is mainly due to the fact that transformers lack the inherent spatial inductive bias of convolutional Neural Networks, which is precisely the key to achieving accurate diagnosis of medical images such as cervical lymph node ultrasound images with rich details and structural information.
\renewcommand{\arraystretch}{1.5}
\begin{table}[ht]
\label{tab2}
\centering
\caption{Precision, Sensitivity, Specificity and F1-S of Different Networks for various diseases.}
\vspace{0.2cm}
\resizebox{1\columnwidth}{!}{
\begin{tabular}{|l|l|l|l|l|}
\hline
\multicolumn{1}{|c|}{\textbf{Model}} & \multicolumn{1}{c|}{\textbf{Type}} & \multicolumn{1}{c|}{\textbf{Precision}} & \multicolumn{1}{c|}{\textbf{Sensitivity}} & \multicolumn{1}{c|}{\textbf{Specificity}} \\ \hline
\multirow{4}{*}{\textbf{SFUSNet(Ours)}}            & Benign                             & \textbf{85.86\%(±3.42)}                          & \textbf{82.53\%(±2.54)}                            & 97.02\%(±0.89)                            \\ \cline{2-5} 
                                     & Normal                             & 99.51\%(±0.48)                          & \textbf{100.00\%(±0.00)}                           & 99.72\%(±0.27)                            \\ \cline{2-5} 
                                     & Primary                            & \textbf{84.93\%(±5.02)}                          & \textbf{84.72\%(±5.15)}                            & 98.83\%(±0.45)                            \\ \cline{2-5} 
                                     & Metastatic                         & \textbf{91.52\%(±1.20)}                          & 92.53\%(±0.85)                            & \textbf{94.40\%(±0.89)}                            \\ \hline
\multirow{4}{*}{DenseNet121}         & Benign                             & 82.25\%(±7.34)                          & 75.87\%(±2.38)                            & 96.31\%(±1.92)                            \\ \cline{2-5} 
                                     & Normal                             & \textbf{99.67\%(±0.65)}                          & 99.75\%(±0.20)                            & \textbf{99.82\%(±0.37)}                            \\ \cline{2-5} 
                                     & Primary                            & 83.88\%(±6.54)                          & 72.87\%(±5.32)                            & 98.92\%(±0.51)                            \\ \cline{2-5} 
                                     & Metastatic                         & 87.36\%(±1.07)                          & 91.99\%(±3.57)                            & 91.33\%(±0.78)                            \\ \hline
\multirow{4}{*}{ResNet101}           & Benign                             & 72.37\%(±4.77)                          & 65.90\%(±3.63)                            & 94.44\%(±1.59)                            \\ \cline{2-5} 
                                     & Normal                             & 99.11\%(±0.97)                          & 99.84\%(±0.20)                            & 99.50\%(±0.55)                            \\ \cline{2-5} 
                                     & Primary                            & 77.96\%(±7.50)                          & 64.45\%(±7.01)                            & 98.57\%(±0.69)                            \\ \cline{2-5} 
                                     & Metastatic                         & 82.61\%(±0.80)                          & 87.51\%(±3.63)                            & 87.98\%(±1.06)                            \\ \hline
\multirow{4}{*}{ConvNext-Large}      & Benign        & 74.96\%(±9.81)     & 30.29\%(±6.26)                          & 97.45\%(±1.75)                            \\ \cline{2-5} 
                                     & Normal                             & 98.94\%(±0.55)                          & 99.67\%(±0.31)                            & 99.40\%(±0.31)                            \\ \cline{2-5} 
                                     & Primary                            & 73.79\%(±14.41)                         & 26.24\%(±10.37)                           & 99.05\%(±0.67)                            \\ \cline{2-5} 
                                     & Metastatic                         & 68.83\%(±2.03)                          & 93.49\%(±3.63)                            & 72.25\%(±3.63)                            \\ \hline
\multirow{4}{*}{SwinT-Small}         & Benign                             & 68.75\%(±2.80)                          & 12.47\%(±1.85)                            & 98.78\%(±0.18)                            \\ \cline{2-5} 
                                     & Normal                             & 98.78\%(±0.57)                          & 99.67\%(±0.31)                            & 99.31\%(±0.33)                            \\ \cline{2-5} 
                                     & Primary                            & 25.71\%(±31.82)                         & 3.79\%(±4.68)                             & 99.84\%(±0.20)                            \\ \cline{2-5} 
                                     & Metastatic                         & 63.80\%(±0.81)                          & \textbf{97.31\%(±0.65)}                            & 64.02\%(±1.32)                            \\ \hline
\multirow{4}{*}{ViT-Base}            & Benign                             & 61.83\%(±6.58)                          & 22.62\%(±3.31)                            & 96.92\%(±0.90)                            \\ \cline{2-5} 
                                     & Normal                             & 98.07\%(±1.27)                          & 99.59\%(±0.26)                            & 98.90\%(±0.73)                            \\ \cline{2-5} 
                                     & Primary                            & 54.73\%(±33.39)                         & 9.78\%(±7.21)                             & 99.40\%(±0.70)                            \\ \cline{2-5} 
                                     & Metastatic                         & 66.08\%(±1.00)                          & 93.42\%(±0.91)                            & 68.74\%(±1.34)                            \\ \hline
\multirow{4}{*}{SwinT V2-Small}      & Benign                             & 27.94\%(±34.38)                         & 2.33\%(±3.55)                             & \textbf{99.75\%(±0.42)}                            \\ \cline{2-5} 
                                     & Normal                             & 96.18\%(±0.87)                          & 99.26\%(±0.55)                            & 97.79\%(±0.52)                            \\ \cline{2-5} 
                                     & Primary                            & 0.00\%(±0.00)                           & 0.00\%(±0.00)                             & \textbf{100.00\%(±0.00)}                           \\ \cline{2-5} 
                                     & Metastatic                         & 61.24\%(±0.56)                          & 96.78\%(±0.62)                            & 60.08\%(±1.09)                            \\ \hline
\end{tabular}}
\end{table}
}
\section{conclusion}
This paper proposed a convolutional neural network based on the spatial and frequency domains for diagnosing diverse cervical lymph node lesions in ultrasound images. Comprehensive evaluation results suggest our network is the state-of-the-art model among various modern architectures. The next step in our research is to address the issue of dataset imbalance. Obviously, due to the existence of class imbalance, the diagnostic ability for Benign lesion and Primary lesion is slightly insufficient. Furthermore, we plan to further test our model in different populations according to gender and age and developing the specific diagnostic systems. Additionally, we also plan to explore the feasibility of conducting human-machine comparative studies, which will help evaluate the model's performance in real-world scenarios with the involvement of human expertise, thus enhancing its overall usefulness and applicability.

\newpage
\bibliographystyle{IEEEbib}
\bibliography{strings,refs}

\end{document}